\documentclass{emulateapj}

\usepackage{txfonts}
\usepackage{times,psfig}

\def\ltsima{$\; \buildrel < \over \sim \;$}
\def\lsim{\lower.5ex\hbox{\ltsima}}
\def\gtsima{$\; \buildrel > \over \sim \;$}
\def\gsim{\lower.5ex\hbox{\gtsima}}

\newcommand{\be}{\begin{equation}}
\newcommand{\en}{\end{equation}}
\newcommand{\ergs}{\rm \ erg \; s^{-1}}

\def\cmdue {\rm \ cm^{-2}}

\def\msole {~M_{\odot}}

\begin{document}
%\received{~~} \accepted{~~}
%\journalid{}{}
%\articleid{}{}

\title{Linking burst-only X--ray binary sources to faint X--ray transients}
%\subtitle

\author{S.~Campana\altaffilmark{1}}

\altaffiltext{1}{INAF-Osservatorio Astronomico di Brera, Via Bianchi 46, I--23807
Merate (Lc), Italy}

\email{sergio.campana@brera.inaf.it}

\begin{abstract}
Burst-only sources are X--ray sources showing up only during short bursts but with no persistent emission (at least 
with the monitoring instrument which led to their discovery). These bursts have spectral characteristics consistent 
with thermonuclear (type I) burst from the neutron star surface, linking burst-only sources to neutron star 
X--ray binary transients. 
We have carried out a series of snapshot observations of the entire sample of burst-only sources with the Swift satellite. 
We found a few sources in outburst and detect faint candidates likely representing their quiescent counterparts.
In addition, we observed three quasi-persistent faint X--ray binary transients.
Finally we discuss burst-only sources and quasi-persistent sources in the framework of neutron star transients.
\end{abstract}


\section{Introduction}

Type I X--ray bursts are observationally characterized by a rapid rise of the X--ray flux 
(of the order of a few seconds) and an exponential decay (usually lasting a few minutes) together with a 
black body spectrum which softens as the burst decays. These bursts are interpreted as thermonuclear
flashes on the surface of a neutron star, thus definitely identifying the
source as a neutron star binary. In the last few years thanks to the long monitoring of the 
Galactic center region with the Wide-Field Cameras onboard BeppoSAX, a number
of sources showing only type I bursts without persistent emission were discovered 
(Cocchi et al. 2001; Cornelisse et al. 2002a, C02a hereafter),  
Upper limits on the quiescent emission were relatively high in the $10^{35}-10^{36}\ergs$ range (C02a). 
This level is significantly lower than the outburst peak luminosity observed in bright X--ray binary 
transients (e.g. Aql X-1, for a review see Campana et al. 1998), thus justifying their name: burst-only sources.
A follow-up of these burst-only sources with short Chandra exposures led to the detection 
of possible quiescent counterparts at a level consistent with other X--ray transients in quiescence, 
i.e. $10^{32}-10^{33}\ergs$ (Cornelisse et al. 2002b, C02b hereafter). 

In addition to classical X--ray binary transients, the increased sensitivity level of monitoring instruments 
led to the discovery of faint X--ray transients with peak
outburst luminosity in the $\sim 10^{36}-10^{37}\ergs$ (Heise et al. 1999;  in't Zand 2001). The most famous 
example is the accreting X--ray millisecond  pulsars (AMXPs)  SAX J1808.4--3658 (as well as many other AMXPs; 
see Wijnands 2005a for a review). Furthermore, Galactic plane X--ray surveys with XMM-Newton and Chandra 
(e.g. Sakano et al. 2005; Muno et al. 2005; Wijnands et al. 2006) revealed the existence of a class of 
very faint X--ray transients (VFXTs) with peak outburst luminosities in the  $10^{34}-10^{36}\ergs$ range. 
A large number of them were found close to the Galactic center (Muno et al. 2005, 2008). 

Close to VFXTs are quasi-persistent sources like 1RXS J170854.4--321857, 1RXS J171824.2--4029 34 
(in't Zand et al. 2005) and XMMU J174716.1--281048 (Del Santo et al. 2007a). These sources are type I bursters, with
the only difference with burst-only sources that they have been detected in a low-level active 
($L\sim 10^{34}-10^{35}\ergs$) state (at least) quite frequently.

In this paper we exploit Swift X--ray Telescope (XRT) observations of six burst-only sources (C02a) as well
as of three quasi-persistent VFXTs (in't Zand et al. 2005; Del Santo et al. 2007a). In Section 2 we describe the general treatment 
of the data. In Section 3 we analyze burst-only Swift XRT data and in Section 4 data of quasi-persistent sources.
In Section 5 we discuss our results.

\section{Swift XRT observations}

The Swift satellite (Gehrels et al. 2004) easily offers the possibility of short snapshot observations thanks to its fast 
repointing capabilities and its low Earth orbit. Burst-only sources are therefore ideal targets to look for signs of activity 
and at the same time collect photons to identify their counterparts in quiescence. In Table 1 (burst-only sources) and 2 
(quasi-persistent VFXTs) we report the log of our observations. 
We concentrate on XRT data (Burrows et al. 2005), being our sources too faint for a detection with the Burst Alert 
Telescope and too absorbed (and faint) for a detection with the UltraViolet/Optical Telescope.
Given the faintness of our sources all the data were collected in Photon Counting mode (PC, providing 2D imaging 
and full spectroscopic resolution).

All data were processed with the standard XRT pipeline within  
HEASOFT 6.5.1 ({\tt xrtpipeline} v.0.12.0) in order to produce screened event
files. PC data were extracted in the 0.3--10 keV energy range with standard grade filtering (0--12). 
Extraction regions were selected depending on source strength. Concentric background regions were 
used to extract background spectra. Appropriate ancillary response files were
generated with the task {\tt xrtmkarf}, accounting for Point Spread Function
losses and CCD defects.  We used v.010 response matrices. Data collected 
after 2007, August 30 were taken with a 6V CCD substrate voltage and require different filtering 
criteria and (slightly) different response matrices (Godet et al. 2009).

\begin{table}
\caption{Swift XRT observation log of burst-only sources.}
\label{log_burst}
{\footnotesize
\begin{center}
\begin{tabular}{ccc}
Obs. ID.            & Date$^*$   &Expos. time$^+$ \\
                          & (yy-mm-dd)&  (ks)              \\
\hline
SAX J1324.5--6313 &&\\
00035708002 &2006-09-30&    5.2\\
00035708003 &2007-03-21&    9.4\\
00035708004 &2007-03-22&    14.2\\
\hline
SAX J1752.3--3128 &&\\
00035712001 &2007-03-19&    0.5\\
\hline
SAX J1753.5--2349 &&\\
00035713001 &2008-02-18&    7.6\\
00035713002 &2008-10-23&    1.0\\
\hline
SAX J 1806.5--2215 &&\\
00035714001 &2006-10-21&    3.9\\
00035714002 &2007-02-11&    3.9\\
\hline
SAX J1818.7+1424 &&\\
00035709001 &2006-09-01&    5.9\\
00035709002 &2006-09-02&    6.4\\
00035709003 &2007-02-14&    0.7\\
00035709004 &2007-02-25&   10.2\\
\hline
SAX J1828.5--1038 &&\\
00035710001 &2007-02-12&    1.4\\
00035710002 &2007-02-27&    1.3\\
00035710003 &2007-04-04&    9.5\\
00035710004 &2007-11-11&    8.5\\
00035710005 &2008-10-31&    3.6\\
00035710006 &2008-11-05&    0.9\\
00035710007 &2008-11-12&    2.6\\
\hline
SAX J 2224.9+5422 &&\\
00035711001 &2006-09-24&    5.0\\
00035711002 &2006-10-28&   10.9\\
00035711004 &2007-05-20&    5.0\\
00035711005 &2008-03-26&    1.9\\
\hline
Swift J1749.4--2807 &&\\
00213190000 &2006-06-02&   16.7\\
00213190001 &2006-06-04&    1.1\\
00213190002 &2006-06-06&    6.1\\
00213190003 &2006-06-07&    9.7\\
00213190004 &2006-06-08&   10.4\\
00213190005 &2006-06-09&    6.2\\
00213190006 &2006-06-10&    5.3\\
\hline
\end{tabular}
\end{center}
}
\noindent $^*$ All observations were taken in Photon Counting mode. Observations after 2007, Aug. 30 were collected with a 
different CCD substrate voltage. 

\end{table}

\begin{table}
\caption{Swift XRT observation log of quasi-persistent sources.}
\label{log_quasi}
{\footnotesize
\begin{center}
\begin{tabular}{ccc}
Obs. ID.            & Date$^*$   &Expos. time \\
                          & (yy-mm-dd)&  (ks)              \\
\hline
1RXS J170854.4--321857 &&\\
00035715001 &2007-02-28&    7.2 \\
00035715001 &2008-04-29&    8.6 \\
\hline
1RXS J171824.2--402934 &&\\
00035716001 &2006-09-25&    2.4\\
00035716002 &2007-01-28&    3.2\\
00090056001 &2008-04-02&    3.3\\
00090056002 &2008-04-09&    2.9\\
00090056003 &2008-04-16&    2.1\\
00090056004 &2008-04-23&    1.5\\
00090056005 &2008-04-30&    1.6\\
00090056006 &2008-05-07&    2.1\\
00090056007 &2008-05-14&    2.3\\
00090056008 &2008-05-26&    1.4\\
00090056009 &2008-04-28&    2.1\\
00090056010 &2008-06-04&    0.7\\
00090056011 &2008-06-07&    0.5\\
00090056012 &2008-06-11&    1.9\\
00090056013 &2008-06-14&    1.4\\
00090056014 &2008-06-21&    2.1\\
00090056015 &2008-06-25&    2.4\\
00090056016 &2008-06-28&    1.7\\
00090056017 &2008-07-02&    2.1\\
00090056018 &2008-07-05&    2.1\\
00090056019 &2008-07-09&    1.3\\
00090056020 &2008-07-12&    1.9\\
00090056021 &2008-07-19&    2.4\\
00090056022 &2008-07-23&    2.1\\
00090056023 &2008-07-26&    2.3\\
00090056024 &2008-07-30&    2.0\\
00090056025 &2008-08-02&    2.4\\
00090056026 &2008-08-06&    2.0\\
00090056027 &2008-08-13&    2.3\\
00090056028 &2008-08-20&    2.2\\
00090056029 &2008-08-27&    2.2\\
00090056030 &2008-09-03&    3.3\\
00090056031 &2008-09-10&    1.8\\
00090056032 &2008-09-17&    2.0\\
00090056033 &2008-09-24&    2.2\\
00090056034 &2008-10-01&    1.0\\
00090056035 &2008-10-08&    2.2\\
00090056036 &2008-10-15&    0.8\\
00090056037 &2008-10-22&    1.7\\
00090056038 &2008-10-29&    0.5\\
\hline
XMMU J174716.1--281048& & \\
00030938001 &2007-05-13&    4.5\\
00030938002 &2007-05-17&    1.3\\
00030938003 &2007-07-03&    1.1\\
00030938004 &2007-08-08&    2.0\\
00030938005 &2008-04-28&    2.0\\
\hline
\end{tabular}
\end{center}
}
\noindent $^*$ All observations were taken in Photon Counting mode.
Observations after 2007, Aug. 30 were collected with a 
different CCD substrate voltage. 

\end{table}

\section{Burst-only sources}

\subsection{SAX J1324.5--6313}

SAX J1324.5--6313 (hereafter SAXJ1324) was discovered by the BeppoSAX WFC through the detection of one 
type I burst over 18.5 ks of monitoring. The $99\%$ error circle is $1.8'$. The upper limit on the quiescent emission is 
$1\times 10^{-12}\ergs\cmdue$ (0.5--7 keV). From the burst strength an upper 
limit on the distance of 6.2 kpc can be set (C02a). This limit on distance (here and in the following) was derived 
by C02a assuming that the burst peak flux is equal or less than the Eddington limit of $2\times 10^{38}\ergs$. 
Given the uncertainties in the companion composition (e.g. Kuulkers et al. 2003) as well as on the source peak flux 
this distance has to be considered as indicative. Follow-up observations were carried out 
with Chandra for 5.1 ks. Five sources were detected within the error circle (C02b). No firm counterparts were 
established however.

\begin{figure}[htbp]
\begin{center}
\psfig{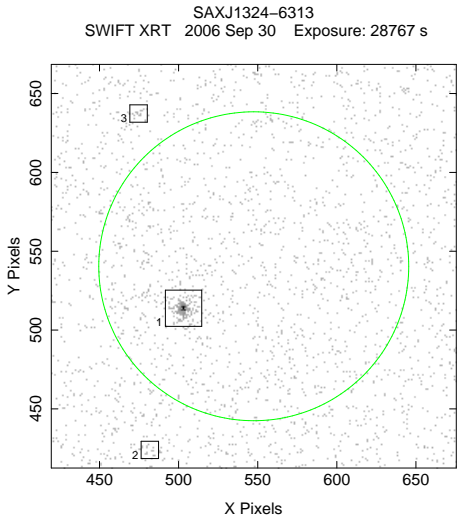}
\end{center}
\caption{SAX J1324.5--6313 field observed by Swift XRT.}
\label{1324_ima}
\end{figure}

We observed SAXJ1324 for a total exposure of 29 ks (splitted into three observations, see Table \ref{log_burst}) 
with the Swift XRT. Within the BeppoSAX error circle we detect only one source (SAXJ1324-1)
coincident with Chandra's source B (error circle $3.7''$ arcsec) in CO2b (see Fig. \ref{1324_ima}). 
Other two sources are detected outside the BeppoSAX WFC error circle. Searching at the known positions of Chandra 
sources we detect one more (weak) source (see Table \ref{1324_source}).

\begin{figure}[htbp]
\begin{center}
\psfig{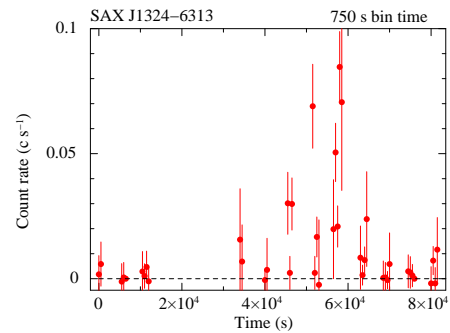}
\end{center}
\caption{ Swift XRT light curve of SAX J1324.5--6313.}
\label{1324_flare}
\end{figure}

We concentrate here on SAXJ1324-1 being the only one detected source within the BeppoSAX WFC error circle.
SAXJ1324-1 was observed to vary among the three Swift XRT exposures at a $\sim 4.6\,\sigma$ level (observed count rate 
are $(6.0\pm1.4)\times 10^{-3}$ c s$^{-1}$, $(6.0\pm1.0)\times 10^{-3}$ c s$^{-1}$ and $(13.3\pm1.2)\times 10^{-3}$ c s$^{-1}$,
respectively, $1\,\sigma$ errors here).  The third observation shows a count rate larger by a factor of two with respect to the 
previous two. In addition, within this last observation signs of flaring activity can be observed (see Fig. \ref{1324_flare}).
No similar variability was observed during the first two observations. 

\begin{table}
\caption{SAX J1324.5--6313 field source detections.}
\label{1324_source}
{\footnotesize
\begin{center}
\begin{tabular}{cccc}
Source name& RA& Dec. & Count rate \\
& (J2000) & (J2000) & (c s$^{-1}$)\\
\hline
SAXJ1324-1 (B)      & 13 24 30.6 &--63 13 45.4&$(9.8\pm0.8)\times10^{-3}$\\
SAXJ1324-2          & 13 24 37.9 &--63 17 18.2&$(5.9\pm2.4)\times10^{-4}$\\
SAXJ1324-3          & 13 24 40.3 &--63 08 55.3&$(7.1\pm2.6)\times10^{-4}$\\
\hline
SAXJ1324-A$^*$      & 13 24 30.2 &--63 12 41& $<7.3\times 10^{-4}$ $3\,\sigma$\\
SAXJ1324-C$^*$      & 13 24 38.0 &--63 12 26& $<1.2\times 10^{-3}$ $3\,\sigma$\\
SAXJ1324-D$^*$      & 13 24 38.3 &--63 13 28& $<1.2\times 10^{-3}$ $3\,\sigma$\\
SAXJ1324-E$^*$      & 13 24 39.4 &--63 13 34& $(4.8\pm2.3)\times10^{-4}$\\
\hline
\end{tabular}
\end{center}
}
\noindent $^*$ Chandra sources with fixed position search in XRT images.
\end{table}

For spectral analysis we extracted separated spectra for observations 1+2 and 3 separately.
We extracted the spectrum of SAXJ1324-1 from a 15 pixel circle centered on source (background was 
extracted from an annulus centered on source with 50 and 70 inner and outer radii, respectively).
We first check if the two spectra are consistent. We fit the two spectra with the same model and free 
normalizations finding good reduced $\chi^2_{\rm red}$ values (in the range 0.8--1). This prompted us to
merge the three sets of data to increase statistics. We then binned the total 266 photons to 10 photons 
per spectral bin and applied Churazov's weighting within XSPEC.
A power law fit provides good results with a reduced $\chi_{\rm red}^2=0.7$ (with 23 degrees of freedom, dof).
The column density is $9.7^{+6.8}_{-4.7}\times 10^{21}\cmdue$ ($90\%$ confidence level throughout the paper). 
This value is consistent with the Galactic column density value of $(1.2-1.5)\times10^{22}\cmdue$ (Kalberla et al. 2005; 
Dickey \& Lockman 1990, respectively).
The power law photon index is $\Gamma=0.8\pm0.4$. A fit with other simple models provide equally good fit (see Table 
\ref{1324_spe}). Given this spectral model the source count rate is larger by a factor of $\sim 3$ with respect 
to the Chandra observation (C02b).

\begin{table}
\caption{SAX J1324--6313: spectral models for SAXJ1324-1.}
\label{1324_spe}
{\footnotesize 
\begin{center}
\begin{tabular}{ccccc}
Model   & Column density       & $\Gamma$ / $kT$        & $\chi^2_{\rm red}$& Flux$^*$\\
                      &($10^{21}\cmdue$)   & -- / (keV)                        & (dof)             & (cgs)\\
\hline
Power law &$9.5^{+7.0}_{-4.5}$  &$0.8^{+0.4}_{-0.4}$       & 0.7 (23)          & $1.4\times 10^{-12}$\\
Black body&$3.1^{+3.9}_{-2.4}$  &$1.7^{+0.4}_{-0.3}$       & 0.9 (23)          & $1.0\times 10^{-12}$\\
Bremsstr. &$14.9^{+5.8}_{-3.9}$ &$>32$                     & 0.9 (23)          & $1.4\times 10^{-12}$\\
NSA       &$4.7^{+4.1}_{-3.1}$  &$1.2^{+0.2}_{-0.1}$       & 0.9 (23)          & $1.0\times 10^{-12}$\\
\hline
\end{tabular}
\end{center}
}
\noindent $^*$ Unabsorbed flux in the 0.5--10 keV energy band.
\end{table}

Chandra's source B was discarded by C02b as a possible counterpart due to its spectral hardness.
We confirm that the spectrum of SAXJ1324-1 is rather hard. Usually the quiescent spectrum of neutron star transients is 
made by a soft component and (often) a hard power law tail. The photon index of this tail is in the 1.5--2 range. 
The soft component is always present apart from a few exceptions: accreting millisecond X--ray pulsars 
(AMXPs, Campana et al. 2004a, 2005; Wijnands et al. 2005a; Heinke et al. 2007), EXO 1745--248 in Terzan 5 
(Wijnands et al. 2005b), as well as for 1H1905+000 for which only very tight upper limits exist (Jonker et al. 2007). 
AMXPs quiescent luminosities are lower with respect to classical transients (e.g. SAX J1808.4--3658 has an unabsorbed 
0.5--10 keV luminosity of $5\times 10^{31}\ergs$). To have this luminosity SAXJ1324 should lie at 0.5 kpc. 
EXO1745-248 has instead a larger quiescent luminosity ($2\times10^{33}\ergs$). To reach a comparable level SAXJ1324 
should lie at 3.4 kpc. In conclusion, despite the hardness of the spectrum we cannot exclude that SAX1324-1 is the 
quiescent counterpart of SAXJ1324. In addition, fast variability (on timescale $\lsim 1000$ s) has also been observed
in X--ray transients in quiescence (Campana et al. 2004b).

\subsection{SAX J1752.3--3128}

SAX J1752.3--3128 (hereafter SAXJ1752) was discovered by the BeppoSAX WFC through the detection of one 
type I burst over 6 Ms (Cocchi et al. 2001). Photospheric radius expansion led to a source of 9.2 kpc. 
Chandra observed this field for 4.7 ks, two sources were detected within the error circle plus a 
brighter one just outside (C02b).

Swift XRT observed SAXJ1752 just once for 0.5 ks (see Table \ref{log_burst}). No sources were detected. 
Upper limits ($3\,\sigma$) of $3\times 10^{-2}$ c s$^{-1}$ at the location of the Chandra sources can 
be set (corresponding, using a Crab-like spectrum and the Galactic absorption, to a Chandra ACIS-S 
rate of $\sim 0.1$ c s$^{-1}$). This indicates that the Chandra observation was much deeper than the Swift one.

\subsection{SAX J1753.5--2349}

SAX J1753.5--2349 (hereafter SAXJ1754) was discovered by the BeppoSAX WFC through the detection of one 
type I burst over 1 Ms of monitoring (in't Zand et al. 1998). The $99\%$ error circle is $2.5'$. From the burst strength an upper 
limit on the distance of 8.8 kpc can be set.  The upper limit on the quiescent flux from the BeppoSAX
WFC is $1.6\times 10^{-12}\ergs\cmdue$ (0.5--7 keV, in't Zand et al. 1998; C02b). Follow-up observations were carried out 
with Chandra for 5.2 ks. No sources were detected. 

\begin{figure}[htbp]
\begin{center}
\psfig{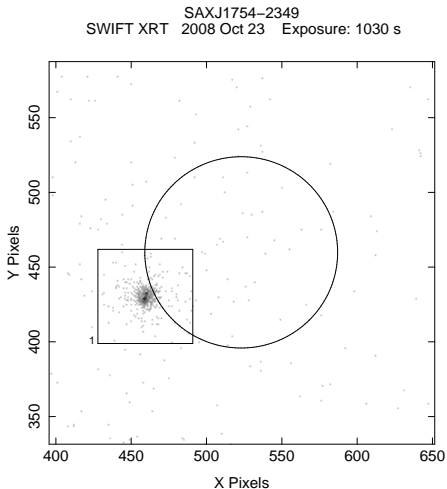}
\end{center}
\caption{SAX J1753.5--2349 field observed by Swift XRT.}
\label{1754_ima}
\end{figure}

SAXJ1754 was observed twice by Swift (see Table \ref{log_burst}). During the first observation (7.6 ks) no sources
were detected within the BeppoSAX WFC error circle with an upper limit of $\sim 1.5\times 10^{-3}$ c s$^{-1}$ ($3\,\sigma$).
Following the detection of an outburst by Swift BAT and RXTE PCA (Markwardt, Krimm \& Swank 2008) and 
INTEGRAL (Cadolle Bel et al. 2008), Swift XRT observed the field for 1 ks.
One source is well visible at the edge of the WFC error circle (see Fig. \ref{1754_ima}; Degenaar \& Wijnands 2008a; 
Starling \& Evans 2008). This source (SAXJ1754-1) is bright at a level of $0.54\pm0.03$ c s$^{-1}$, 
i.e. $\sim 350$ times brighter than in the previous observation. The source is constant within the observation.

Due to a small trim (3 pixels) across the three orbits comprising the second observation, we extracted photons on an orbit
by orbit basis. For each orbit we used a 30 pixel circular region for the source and an annular region  with inner and outer radii 
of 60 and 80 pixels, respectively, for the background. Photons were binned to 20 counts per spectral bin. 
We then generated single arf files and sum them together. The spectrum is rather hard. Good fit with a  single 
component spectral model can be obtained only with a power law or a bremsstrahlung with an absorption higher than 
the Galactic value ($8.6\times10^{21}\cmdue$, see Table \ref{1754_spe}). Fixing the column density to the Galactic value 
double component models encounter problems too. A black body plus power law model can fit the data only for negative values of $\Gamma$.
A black body plus a Comptonization model (COMPTT) provides instead good results, with even too much freedom:
if we fix the plasma optical depth $\tau=10$ we cannot constrain the plasma temperature. The fit is good 
($\chi^2_{\rm red}=0.9$, 25 dof). The 0.5--10 keV unabsorbed flux is in line with single component models 
($5.4\times 10^{-11}\ergs \cmdue$), resulting in a peak luminosity $5\times10^{35}\ergs$ and a quiescent upper limit 
of $10^{33}\ergs$. 

\begin{table}
\caption{SAX J1753.5--2349: spectral models for SAXJ1754-1.}
\label{1754_spe}
{\footnotesize 
\begin{center}
\begin{tabular}{ccccc}
Model   & Column density       & $\Gamma$ / $kT$        & $\chi^2_{\rm red}$& Flux$^*$\\
                      &($10^{21}\cmdue$)     & -- / (keV)             & (dof)             & (cgs)\\
\hline
Power law  &$18.9^{+5.0}_{-6.3}$  &$2.0^{+0.4}_{-0.4}$     & 1.0 (26)          & $8.3\times 10^{-11}$\\
Power law  &8.6 (fix)             &$1.2^{+0.1}_{-0.1}$     & 1.5 (27)          & $5.9\times 10^{-11}$\\
Black body &$5.7^{+2.9}_{-2.2}$   &$1.1^{+0.1}_{-0.1}$     & 1.6 (26)          & $3.8\times 10^{-11}$\\
Bremsstr.  &$15.0^{+4.3}_{-3.2}$  &$8.1^{+10.6}_{-3.5}$    & 1.0 (26)          & $6.5\times 10^{-11}$\\
Bremsstr.  &8.6 (fix)             &$>25$                   & 1.5 (26)          & $5.8\times 10^{-11}$\\
NSA        &$22.6^{+3.3}_{-2.8}$  &$0.5^{+0.1}_{-0.1}$     & 2.9 (26)          & $4.3\times 10^{-11}$\\
\hline
\end{tabular}
\end{center}
}
\noindent $^*$ Unabsorbed flux in the 0.5--10 keV energy band.
\end{table}

\subsection{SAX J1806.5--2215}

SAX J1806.5--2215 (hereafter SAXJ1807) was discovered by the BeppoSAX WFC through the detection of two 
type I burst over 1 Ms of monitoring (in't Zand et al. 1998). The $99\%$ error circle is $2.9'$. From the burst strength an upper 
limit on the distance of 8.0 kpc can be set. The upper limit on the quiescent flux from the BeppoSAX
WFC is $1.5\times 10^{-12}\ergs\cmdue$ (0.5--7 keV, in't Zand et al. 1998; C02b). 
The RXTE ASM light curve of SAX J1806.5--2215 showed a detection during the period of the occurrence of the X--ray bursts 
observed with the BeppoSAX WFC. The maximum persistent flux was $2\times 10^{-10}\ergs\cmdue$ (2--10 keV), 
and it slowly decreased over time. At a distance of 8 kpc, this corresponds to a peak luminosity of $2\times 
10^{36}\ergs$ (C02b). This implies that SAXJ1807 is a faint X--ray transient.
Follow-up observations were carried out with Chandra for 4.8 ks. Nine sources were detected. Of these four are outside 
the BeppoSAX WFC error circle, one is hard and one has a too bright optical counterpart (C02b). 

Swift XRT observed the field of SAXJ1807 twice for a total observing time of 7.9 ks. Two sources were detected, 
all outside the BeppoSAX WFC error circle. None of the Chandra sources were detected (see Table \ref{1807_source}).
Chandra's source B is the brightest one. Given its count rate and assuming a Crab like spectrum absorbed
at the Galactic value ($N_H=1.2\times 10^{22}\cmdue$) we can predict a count rate of $\sim 2\times 10^{-3}$ c s$^{-1}$ 
for the Swift observation,
consistent with our upper limit (Table  \ref{1807_source}). Sources SAXJ1807-1 and SAXJ1807-2 lie at $4.5'$ and $3.5'$ 
from the BeppoSAX position.
SAXJ1807-1 was also detected by Chandra at a rate consistent with the Swift detection, on the contrary SAXJ1807-2 was not
detected by Chandra. With the same spectral parameters as before we can estimate a flux increase by a factor of $\sim 7$.
Despite its large offset this large flux increase might be interesting. We extracted photons from a 8 pixel circular region
totaling 14 counts (background is extracted from a concentric annular region of 40 and 60 inner and outer radii, respectively
and accounts for $15\%$ of the counts). Using Cash statistics we can estimate a power law photon index of $\Gamma=3.5\pm1.1$
or a black body temperature $T=0.6^{+0.3}_{-0.2}$ keV (fixing the column density to the Galactic value). 
The 0.5--10 keV unabsorbed flux is in the $2-6\times10^{-13}\ergs\cmdue$. At a distance of 8 kpc this corresponds to 
$1-4\times 10^{33}\ergs$.

\begin{table}
\caption{SAX J1806.5--2215 field source detections.}
\label{1807_source}
{\footnotesize
\begin{center}
\begin{tabular}{cccc}
Source name& RA& Dec. & Count rate \\
& (J2000) & (J2000) & (c s$^{-1}$)\\
\hline
SAXJ1807-1          & 18 06 40.1&--22 19 25.6&$(3.3\pm0.8)\times10^{-3}$\\
SAXJ1807-2          & 18 06 48.5&--22 14 00.8&$(2.0\pm0.7)\times10^{-3}$\\
\hline
SAXJ1807-A$^*$      & 18 06 18.1&--22 15 39  & $<1.8\times 10^{-3}$ $3\,\sigma$\\
SAXJ1807-B$^*$      & 18 06 18.5&--22 17 24  & $<2.3\times 10^{-3}$ $3\,\sigma$\\
SAXJ1807-C$^*$      & 18 06 19.9&--22 18 03  & $<1.5\times 10^{-3}$ $3\,\sigma$\\
SAXJ1807-D$^*$      & 18 06 31.7&--22 13 19  & $<1.8\times 10^{-3}$ $3\,\sigma$\\
SAXJ1807-E$^*$      & 18 06 35.8&--22 15 01  & $<2.3\times 10^{-3}$ $3\,\sigma$\\
SAXJ1807-F$^*$      & 18 06 36.8&--22 15 26  & $<3.0\times 10^{-3}$ $3\,\sigma$\\
SAXJ1807-G$^*$      & 18 06 37.4&--22 17 22  & $<2.6\times 10^{-3}$ $3\,\sigma$\\
SAXJ1807-H$^*$      & 18 06 43.6&--22 16 06  & $<1.6\times 10^{-3}$ $3\,\sigma$\\
SAXJ1807-I$^*$      & 18 06 43.8&--22 18 42  & $<1.4\times 10^{-3}$ $3\,\sigma$\\
\hline
\end{tabular}
\end{center}
}
\noindent $^*$ {Chandra sources with fixed position search in XRT images}.
\end{table}

\subsection{SAX J1818.7+1424}

SAX J1819.7+1424 (hereafter SAXJ1819) was discovered by the BeppoSAX WFC through the detection of one 
type I burst over 1.6 Ms of monitoring (C02a). The $99\%$ error circle is $2.9'$. From the burst strength an upper 
limit on the distance of 9.4 kpc can be set.  The upper limit on the quiescent flux from the BeppoSAX
WFC is $5.5\times 10^{-12}\ergs\cmdue$ (C02b). Follow-up observations were carried out 
with Chandra for 4.8 ks. Eight sources were detected. Of these three are outside the BeppoSAX WFC error circle and two 
have a too bright optical counterpart (C02b). 

\begin{figure}[htbp]
\begin{center}
\psfig{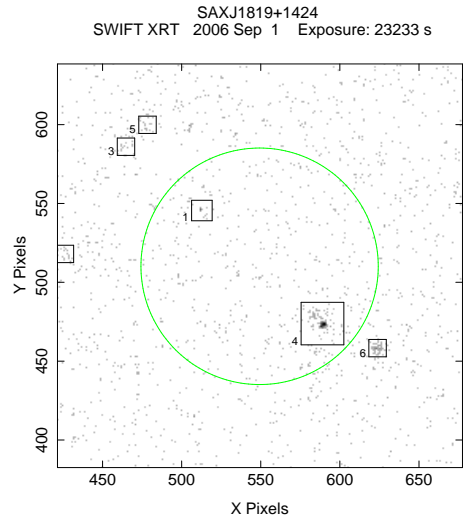}
\end{center}
\caption{SAX J1819.7+1424 field observed by Swift XRT.}
\label{1819_ima}
\end{figure}

Swift observed SAX1819 four times. We detected six sources, two within the BeppoSAX WFC error circle (SAX1819-1 and 
SAX1819-4, see Table \ref{1819_source} and Fig. \ref{1819_ima}). Source SAX1819-4 (source D in Chandra) is consistent 
with the bright star ($V=7.6$) HD168344 and was already discarded by C02b. The high source flux is also due to 
some optical flux leaking through the XRT filter.

\begin{table}
\caption{SAX J1818.7+1422 field source detections.}
\label{1819_source}
{\footnotesize
\begin{center}
\begin{tabular}{cccc}
Source name& RA& Dec. & Count rate \\
& (J2000) & (J2000) & (c s$^{-1}$)\\
\hline
SAX J1819-1          & 18 18 49.9 &+14 25 35.3&$(1.01\pm0.28)\times 10^{-3}$\\
SAX J1819-2          & 18 19 03.9 &+14 24 30.4&$(1.06\pm0.29)\times 10^{-3}$\\
SAX J1819-3          & 18 18 57.7 &+14 27 10.7&$(7.86\pm2.54)\times 10^{-4}$\\
SAX J1819-4 (D)      & 18 18 37.5 &+14 22 46.3&$(1.48\pm0.10)\times 10^{-2}$\\
SAX J1819-5 (H)      & 18 18 55.4 &+14 27 43.5&$(7.64\pm2.21)\times 10^{-4}$\\
SAX J1819-6 (A)      & 18 18 31.9 &+14 22 09.6&$(3.92\pm0.56)\times 10^{-3}$\\
\hline
SAX J1819-B$^*$      & 18 18 34.3 &+14 26 29 &$<9.2\times 10^{-4}$ $3\,\sigma$\\
SAX J1819-C$^*$      & 18 18 35.6 &+14 22 32 &$<1.4\times 10^{-3}$ $3\,\sigma$\\
SAX J1819-E$^*$      & 18 18 37.8 &+14 22 06 &$(5.70\pm2.58)\times 10^{-4}$\\
SAX J1819-F$^*$      & 18 18 38.6 &+14 22 59 &$(1.65\pm0.36)\times 10^{-3}$\\
SAX J1819-G$^*$      & 18 18 48.3 &+14 22 43 &$(5.21\pm2.19)\times 10^{-4}$\\
\hline
\end{tabular}
\end{center}
}
\noindent $^*$ {Chandra sources with fixed position search in XRT images}.
\end{table}

We carry out spectral analysis for sources SAX1819-6 and SAX1819-F (see Table  \ref{1819_source}). We extracted photons 
from a circular region with 10 and 5 pixel radii, respectively (the small radius for source F was dictated by the
closeness of source SAX1819-4). Background photons were extracted from an annular region of 40 (60) inner (outer) radius,
centered on source. We extracted 78 and 27 counts, respectively. Given the low number of counts we grouped 
photons by 5 and use Churazov's weighting scheme within XSPEC. 

For both sources the spectrum is soft and it cannot be fit with the full Galactic absorption ($1.0\times10^{22}\cmdue$). 
The spectrum of SAX1819-6 spectrum is the softest among the two. It can be fit equally well by any of the considered 
single component models (see Table \ref{1819_spe}), even if the power law and the bremsstrahlung models describe 
better the high energy part of the data. This is also the case for SAX1819-F. In the case of a power law fit and at the 
maximum source distance the (upper limit) on source luminosities are $<1\times 10^{32}\ergs$ and $<7\times 10^{32}\ergs$,
respectively. Based on these spectral fit we cannot establish which is the true counterpart of SAXJ1819, if any.

\begin{table}
\caption{SAX J1818.7+1424: spectral models for sources SAX1819-6 and SAX1819-F.}
\label{1819_spe}
{\footnotesize 
\begin{center}
\begin{tabular}{ccccc}
SAXJ1819-6     & Column density     & $\Gamma$ / $kT$       & $\chi^2_{\rm red}$& Flux$^*$\\
Model &($10^{21}\cmdue$)   & -- / (keV))           & (dof)             & (cgs)\\
\hline
Power law &$2.2^{+2.6}_{-0.7}$ &$3.2^{+1.7}_{-1.0}$    & 0.5 (13)          & $1.5\times 10^{-13}$\\
Black body&$<1.4$              &$0.3^{+0.1}_{-0.1}$    & 0.8 (13)          & $6.2\times 10^{-14}$\\
Bremsstr. &$<1.9$              &$1.0^{+1.0}_{-0.5}$    & 0.6 (13)          & $8.9\times 10^{-14}$\\
NSA       &$1.3^{+1.2}_{-0.7}$ &$0.11^{+0.01}_{-0.01}$ & 0.8 (13)          & $9.2\times 10^{-14}$\\
\hline
SAXJ1819-F     & Column density     & $\Gamma$ / $kT$       & $\chi^2_{\rm red}$& Flux$^*$\\
Model &($10^{21}\cmdue$)   & -- / (keV))           & (dof)             & (cgs)\\
\hline
Power law &$<3.7$              &$1.9^{+2.3}_{-0.8}$    & 1.2 (2)           & $6.9\times 10^{-14}$\\
Black body&$<2.6$              &$0.4^{+0.2}_{-0.2}$    & 2.4 (2)           & $3.9\times 10^{-14}$\\
Bremsstr. &$<2.2$              &$>0.8$                 & 1.2 (2)           & $5.6\times 10^{-14}$\\
NSA       &$1.1^{+21}_{-1.0}$  &$0.09^{+0.03}_{-0.04}$ & 3.7 (2)           & $4.0\times 10^{-14}$\\
\hline
\end{tabular}
\end{center}
}
\noindent $^*$ Unabsorbed flux in the 0.5--10 keV energy band.
\end{table}

\subsection{SAX J1828.5--1037}

SAX J1828.5--1037 (hereafter SAXJ1828) was discovered by the BeppoSAX WFC through the detection of one 
type I burst (C02a). The $99\%$ error circle is $2.8'$. From the burst strength an upper 
limit on the distance of 6.2 kpc can be set. SAXJ1828 was previously detected during a ROSAT observation with a 
0.5--2.5 keV flux of $2\times10^{-12}\ergs\cmdue$ (C02b). SAX J1828 was also detected by XMM-Newton during a scan 
of the Galactic plane (Hands et al. 2004). SAXJ1828 was found in a bright state with an unabsorbed 0.5--10 keV 
flux of $10^{-11}\ergs\cmdue$. At a distance of 6.2 kpc this implies a luminosity of $5\times10^{34}\ergs$.
The spectrum was fit with a thermal bremsstrahlung model with equivalent temperature of 7 keV and a column density of 
$N_H=5\times 10^{22}\cmdue$ (much larger than the Galactic value of $N_H=1.9\times10^{22}\cmdue$). 

Swift XRT observed SAXJ1828 for 20.7 ks. Given the detection of a transient source by XMM-Newton we focus on 
this much smaller error circle. A faint source is barely visible at the XMM-Newton position (there are other five sources
in the close vicinity of the BeppoSAX WFC error circle). SAXJ1828 is detected with a count rate of 
$(7.3\pm2.6)\times 10^{-4}$ c s$^{-1}$. Assuming the same spectrum and absorption as in the XMM-Newton observation
we estimate a 0.5--10 keV unabsorbed flux of $1.5\times10^{-13}\ergs\cmdue$ ($1.9\times10^{-13}\ergs\cmdue$ in the
case of a power law with photon index $\Gamma=2$). This highlights a flux decrease by a factor of $\sim 80$ confirming the transient
nature of the counterpart. This flux correspond to a quiescent luminosity of $\sim 7\times 10^{32}\ergs$.

A new outburst from this source was detected at the end of Oct. 2008 (Degenaar \& Wijnands 2008b). 
Three observations were carried out by Swift XRT (see Table \ref{log_burst}). We extracted spectra from the same 15 
pixel region and the background from an annular region 40 (60) pixel inner (outer) radius.
We bin the spectra of the three observations containing 141, 40 and 261 photons, respectively, to 10 counts 
per spectral bin and applied Churazov's weighting scheme. We then fit the three spectra keeping the same column density. 
We find a heavily absorbed source with $N_H=4.1^{+1.5}_{-0.6}\times 10^{22}\cmdue$, consistent with the XMM-Newton value.
The power law photon indeces are $1.7^{+0.7}_{-0.6}$, $1.4^{+1.0}_{-0.9}$ and $1.7^{+0.6}_{-0.5}$, respectively. 
The overall $\chi^2_{\rm red}=0.7$ 
with 36 dof. A fit with the same power law photon index provides similar results ($\chi^2_{\rm red}=0.7$ with 38 dof) and
a photon index $\Gamma=1.6^{+0.6}_{-0.4}$. The 0.5--10 keV unabsorbed fluxes for the three observations are 9.2, 9.7 and $22.9\times 
10^{-11}\ergs\cmdue$, resulting in luminosities of 4.5, 4.7 and $1.1\times 10^{36}\ergs$, respectively. Given the upper limit on 
the source distance, these luminosities are upper limits too.

\subsection{SAX J2224.9+5421}

SAX J2224.9+5421 (hereafter SAXJ2225) was discovered by the BeppoSAX WFC through the detection of one 
short (2.6 s) type I burst (C02a). The fact that the type I burst is short casts some doubts on the X--ray binary nature 
of the object. At the same time the goodness of the black body fit to the burst spectrum and the absence of 
such short X--ray flashes hints for a Galactic object.
The $99\%$ error circle is $3.2'$. From the burst strength an upper 
limit on the distance of 7.1 kpc can be set.  A strong upper limit on the flux was obtained a few hours after the burst 
with a repointing of the BeppoSAX satellite. The source was not detected with the  MECS instrument  with an upper 
limit on the 2--10 keV flux of $1.3\times 10^{-13}\ergs\cmdue$ (C02b), which translates to an upper limit in luminosity
of $\lsim 8\times 10^{32}\ergs$. This source was not covered by Chandra observations.

\begin{table}
\caption{SAX J2224.9+5421 field source detections.}
\label{2224_source}
{\footnotesize
\begin{center}
\begin{tabular}{ccccc}
Source name& RA& Dec. & Count rate \\
& (J2000) & (J2000) & (c s$^{-1}$)\\
\hline
SAXJ2225-1          & 22 24 53.1 &+54 22 36.7 &$(1.47\pm0.34)\times 10^{-3}$\\
SAXJ2225-2          & 22 24 41.4 &+54 24 51.5 &$(1.02\pm0.31)\times 10^{-3}$\\
SAXJ2225-3          & 22 24 25.3 &+54 20 59.3 &$(7.15\pm2.73)\times 10^{-4}$\\
SAXJ2225-4          & 22 24 18.8 &+54 19 49.2 &$(1.02\pm0.31)\times 10^{-3}$\\
\hline
\end{tabular}
\end{center}
}
\end{table}

\begin{figure}[htbp]
\begin{center}
\psfig{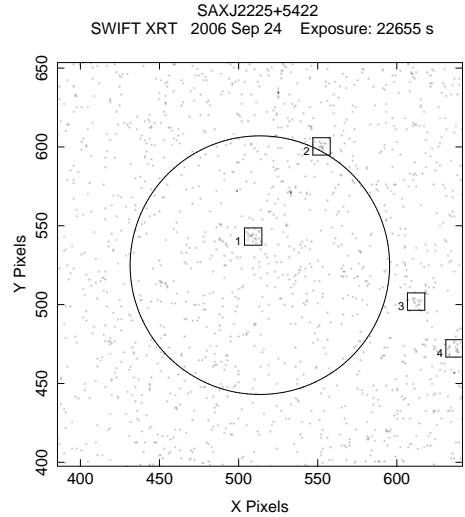}
\end{center}
\caption{SAX J2224.9+5421 field observed by Swift XRT.}
\label{2225_ima}
\end{figure}

Swift XRT observed SAXJ2225 four times for a total exposure time of 22.7 ks. There is just one source within the BeppoSAX
WFC error box and one close to it (see Fig. \ref{2225_ima} and Table \ref{2224_source}).
For SAXJ2225-1 we collected just 24 photons in the 0.3--10 keV energy range. Data were rebinned to 6 photons per channel 
and Churazov's weighting scheme was used. The column density was fixed
to the Galactic value of $5\times 10^{21}\cmdue$ (leaving free the column density it will converge to a value higher
than the Galactic value). The source spectrum is hard. At the maximum distance of 7.1 kpc the X--ray luminosity is 
$8\times 10^{32}\ergs$. In the 2--10 keV the upper limit in luminosity is  $7\times10^{32}\ergs$, i.e. comparable to 
the BeppoSAX MECS upper limit.

\begin{table}
\caption{SAX J2224.9+5422: spectral models for SAXJ2225-1.}
\label{2224_spe}
{\footnotesize 
\begin{center}
\begin{tabular}{ccccc}
Model   & Column density    & $\Gamma$ / $kT$   & $\chi^2_{\rm red}$& Flux$^*$\\
           &($10^{21}\cmdue$)  & -- / (keV))       & (dof)             & (cgs)\\
\hline
Power law  &0.5 (fix)          &$1.1^{+0.8}_{-1.1}$& 0.6 (2)           & $1.3\times 10^{-13}$\\
Black body &0.5 (fix)          &$0.9^{+1.2}_{-0.4}$& 0.7 (2)           & $7.4\times 10^{-14}$\\
Bremsstr.  &0.5 (fix)          &$>3.9$             & 0.7 (2)           & $1.1\times 10^{-13}$\\
\hline
\end{tabular}
\end{center}
}
\noindent $^*$ Unabsorbed flux in the 0.5--10 keV energy band.
\end{table}

\subsection{Swift J1749.4--2807}

\begin{figure}[htbp]
\begin{center}
\psfig{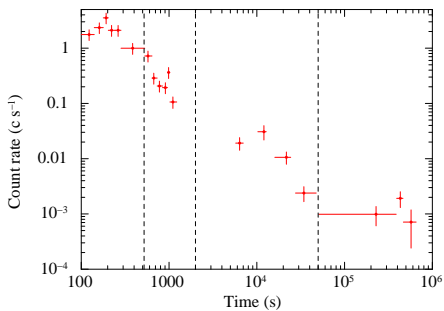}
\end{center}
\caption{Light curve of Swift J1749.4--2807 as observed by Swift XRT. The three vertical dashed lines indicate the three 
time intervals in which spectra were extracted.}
\label{1749_lc}
\end{figure}

One additional burst-only source was detected by Swift in 2006. 
Swift J1749.4--2807 (Swift J1749 in the following) showed a soft burst detected only up to $\sim 40$ keV with the Burst Alert 
Telescope (Schady et al. 2006). A black body fit provided a temperature of 3 keV. An upper limit on distance of $6.7\pm1.3$ 
kpc can be set. The source was then followed up from 83 s to $\sim 8$ d by the XRT telescope (Wjinands et al. 2009).
The source showed a monotonic decrease following a power law with index $-0.99\pm0.05$. Swift J1749 decreased in flux by 
three orders of magnitude. The unabsorbed 0.5--10 keV luminosity in outburst resulted in $\sim 10^{36}\ergs$ making this 
source a likely faint X--ray transient (Wjinands et al. 2009).
We reanalyzed the spectra from this source. We took three spectra in the three intervals described in Fig. \ref{1749_lc},
avoiding the first part during which the source rate was piled-up. 
The main aim was to derive a flux to count rate conversion in the lowest count rate part. 
We then fit together the three spectra keeping fixed the column density and allow a power law model to vary.
We confirm that the first spectrum is soft with $\Gamma=2.7^{+1.5}_{-1.1}$ and highly absorbed ($N_H=4.0^{+3.1}_{-1.9}\times 
10^{22}\cmdue$). The second spectrum is harder with $\Gamma=2.2^{+1.4}_{-1.0}$ and the third even harder $\Gamma=0.5\pm1.3$
(the overall fit is good with $\chi^2_{\rm red}=1.0$ with 15 dof). Based on this spectral parametrization, the latest point 
in the light curve presented by Wjinands et al. (2009) corresponds to a 0.5--10 keV unabsorbed flux of $(5\pm4)\times10^{-14}\ergs\cmdue$
(whereas the model extrapolation corresponds to $2\times10^{-14}\ergs\cmdue$) and in turn to a luminosity of $(3\pm2)\times10^{32}\ergs$
($10^{32}\ergs$). This level coincides within the uncertainties with what observed during a serendipitous 
observation of the same field carried out by XMM-Newton 6 years before (Halpern 2006). 

\section{Quasi-persistent sources}

\subsection{1RXS J170854.4--321857}

1RXS J170854.4--321857 (RXSJ1708 in the following) is a ROSAT all sky-survey source from which a 
type I burst was detected by the BeppoSAX WFC, without any associated persistent emission 
(in't Zand et al. 2005). The burst lasted about 10 min and showed signs of photospheric radius expansion,
leading to a source distance estimate of $13\pm2$ kpc. The $99\%$ confidence error radius is $1.8'$ containing
the ROSAT source. RXSJ1708 has been continuously detected through the RXTE ASM 
and it has also been detected by INTEGRAL (in't Zand et al. 2005). Chandra observed the source constraining the 
power law spectrum to $\Gamma=1.9\pm0.2$ and $N_H=(4.0\pm1.0)\times 10^{21}\cmdue$. 
in't Zand et al. (2005) concluded that the source is variable but it remains in a low-activity state around 
$(2-4)\times 10^{36}\ergs$ (for a 13 kpc distance). 

Swift observed twice the source. In the first observation the source is heavily piled-up and an annular region with 
4 (40) pixels of inner (outer) radius was considered. In the second observation the source is much fainter and 
we extracted photons from a 25 pixel radius circular region. Background was extracted from an annular region 
with 60 and 80 inner and outer radii, respectively. 
We fit together the two spectra keeping the same column density. 
The first spectrum is consistent with the Chandra one with $\Gamma=1.98\pm0.05$ and the column density is consistent 
too $N_H=(4.3\pm0.2)\times 10^{21}\cmdue$ ($90\%$ confidence level). The source 0.5--10 keV luminosity during the 
first observation is $3.3\times 10^{36}\ergs$, at the high end of the observed range. 
The second observation found the source in a much fainter state ($3.6\times10^{35}\ergs$). 
The spectral index is also softer $\Gamma=2.4\pm0.1$. This is the faintest level ever attained by RXSJ1708. 

\subsection{1RXS J171824.2--402934}

1RXS J171824.2--402934 (RXSJ1718 in the following) was detected and identified during the ROSAT all sky survey (Sep. 1990), 
and it was later reobserved twice with the ROSAT HRI (Mar. and Sep. 1994; Kaptein et al. 2000). The BeppoSAX WFC never 
detected the persistent emission but a bright burst. An upper limit on distance is in the 6--8 kpc range (Kaptein et al. 2000).
In the following we assume $d=6.$ kpc. RXSJ1718 was observed by Chandra for 15 ks. The source is well detected with a 
power law spectrum $\Gamma=2.1\pm0.2$ ($N_H=1.3^{+0.2}_{-0.1}\times10^{22}\cmdue$). The observed 0.5--10 keV 
luminosity is $5\times 10^{34}\ergs$ (in't Zand et al. 2005).

\begin{figure}[htbp]
\begin{center}
\psfig{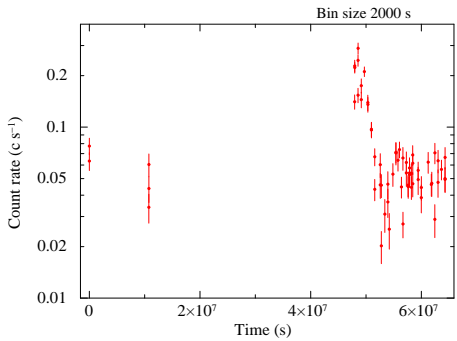}
\end{center}
\caption{1RXS J171824.2--402934  light curve observed by Swift XRT.}
\label{rxs1718_lc}
\end{figure}

Swift observed RXSJ1718 for 78 ks. Two short observations were carried out in Sep. 2006 and Jan. 2007, whereas the 
bulk of the data comes from prolonged observation following a source flare (see Table \ref{log_quasi} and Fig. 
\ref{rxs1718_lc}). The source underwent a rebrightening by a factor of $\sim 10$, showing strong variability among different
observations (Fig. \ref{rxs1718_lc}). Given the large wealth of data we divided the entire set. Data before 2008 were grouped together.
The flare data were then divided according to the source count rate into four intervals: $>0.2$ c s$^{-1}$; $0.1-0.2$ c s$^{-1}$; 
$0.05-0.1$ c s$^{-1}$ and $<0.05$ c s$^{-1}$. Data were extracted from a circular region centered on source with radius 20, 25, 20,
20 and 15 pixels, respectively. Background was extracted from an annular region with 60 (80) inner (outer) radius.

An absorbed power law model can fit all the data. Spectral parameters and fluxes are reported in Table \ref{rxs1718_spe}.
The overall fit with variable $\Gamma$ but the same column density ($N_H=(1.2\pm0.1)\times 10^{22}\cmdue$) is good with
$\chi^2_{\rm red}=0.99$ for 248 dof. As can be seen from Table \ref{rxs1718_spe} there is a clear softening of the 
spectrum at decreasing luminosities. This is opposite to what is usually observed in X--ray binary transients where
a spectral hardening is observed at lower X--ray luminosities. An equivalent fit can be obtained with a power law model 
(with the same $\Gamma$ for al the spectra) plus the same black body component for all the spectra leaving free
only the normalization of the power law component across different observations. The fit is even better 
than the previous one with $\chi^2_{\rm red}=0.91$ for 250 dof). The best fit values are $N_H=(1.3\pm0.1)\times 10^{22}\cmdue$,
$k\,T=0.27\pm0.04$ keV, $R=2.2^{+1.5}_{-0.8}$ km and $\Gamma=2.0\pm0.1$. The data are therefore consistent with a 
progressive decrease of the power law component only with luminosity.

\begin{table}
\caption{1RXS J171824.2--402934: spectral fits.}
\label{rxs1718_spe}
{\footnotesize 
\begin{center}
\begin{tabular}{cccc}
Interval              &Exposure time& $\Gamma$           & Luminosity$^*$\\
                      & (ks)        &                    & (cgs)\\
\hline
Before 2008           &5.6          &$2.3^{+0.3}_{-0.3}$ & $3.9\times 10^{34}$\\
$>0.2$ c s$^{-1}$     &6.0          &$1.9^{+0.1}_{-0.1}$ & $1.6\times 10^{35}$\\
$0.1-0.2$ c s$^{-1}$  &7.0          &$2.1^{+0.1}_{-0.1}$ & $9.2\times 10^{34}$\\
$0.05-0.1$ c s$^{-1}$ &32.1         &$2.4^{+0.1}_{-0.1}$ & $4.3\times 10^{34}$\\
$<0.05$ c s$^{-1}$    &27.2         &$2.4^{+0.2}_{-0.2}$ & $2.5\times 10^{34}$\\
\hline
\end{tabular}
\end{center}
}
\noindent $^*$ Unabsorbed flux in the 0.5--10 keV energy band.
\end{table}

\subsection{XMMU J174716.1--281048}

XMMU J174716.1--281048 (XMMJ1747 in the following) was discovered as a faint X--ray 
transient in 2003 with XMM-Newton (Sidoli \& Mereghetti 2003). Exposure of the same region with 
XMM-Newton and Chandra did not reveal any source down to a limiting 2--10 keV flux of $\sim 3\times 
10^{-14}\ergs\cmdue$ (Del Santo et al. 2007a). The 2--10 keV flux at the discover was $7\times10^{-12}\ergs\cmdue$.
The spectrum was fit well by an absorbed power-law with $\Gamma=2.1\pm0.1$ and a high column density of 
$N_H=(8.9\pm0.5)\times10^{22}\cmdue$. A further XMM-Newton observation in 2005 detected the source with a similar
spectrum and a 2-10 keV flux of $3\times10^{-12}\ergs\cmdue$ (Del Santo et al. 2007a). 

INTEGRAL detected several type I bursts from this source, leading to a distance estimate of $\sim 8$ kpc,
consistent with lying close to the Galactic center (Del Santo et al. 2007a; Del Santo et al. 2007b).
The source was always found in an active state, testifying for the long duration of the outburst. 

\begin{figure}[htbp]
\begin{center}
\psfig{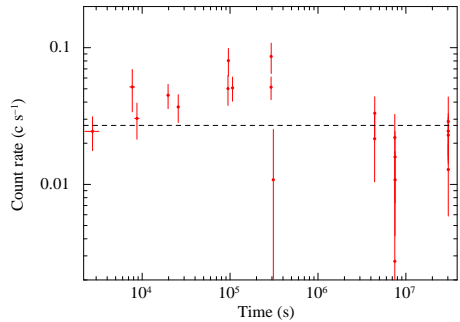}
\end{center}
\caption{XMMU J174716.1--281048  light curve observed by Swift XRT.}
\label{xmmj1747_lc}
\end{figure}

Swift observed several times XMMJ1747 for a total time of 10.9 ks (see also Degenaar \& Wijnands 2007; Sidoli et al. 2008a; 
Sidoli et al. 2008b). The source is highly variable by more than a 
factor of 10 (Fig. \ref{xmmj1747_lc}). We first divided the data into intensity bins (above and below a rate of 0.04 c s$^{-1}$) 
but found no spectral differences (including errors) so we merge the entire dataset into a single spectrum.
Data were rebin to 20 photons per energy channel. The (mean) spectrum can be well fit ($\chi^2_{\rm red}=0.8$ for 18 dof)
with an absorbed $N_H=(7.9\pm2.9)\times 10^{22}\cmdue$ power law with photon index $\Gamma=1.8\pm0.8$. 
Fixing the column density to the XMM-Newton value we have $\Gamma=1.9\pm0.3$. The mean flux is $8.8\times 
10^{-12}\ergs\cmdue$ (the conversion from observed rate to unabsorbed 2--10 keV flux is $2.8\times 10^{-10}\ergs\cmdue$). 
According to this spectrum XMMJ1747 spanned a 2--10 keV luminosity of $2\times 10^{34}-2\times 10^{35}\ergs$.

%\begin{figure*}[htbp]
%\begin{center}
%\psfig{figure=1808_lc.ps,width=12truecm,angle=-90}
%\end{center}
%\caption{SAX J1808 light curve observed with Swift XRT during the 2005 outburst.}
%\label{lc}
%\end{figure*}

\section{Discussion}

In the framework of low mass transients, burst-only sources stand as peculiar, firmly established, neutron star binary systems.
The sources were discovered by long monitoring of the Galactic center region by the WFC onboard BeppoSAX, 
through type I burst activity but without the detection of persistent emission (C02a). 
This behavior has been recently interpreted in terms of sedimentation of CNO nuclei, leading to strong H/He flashes
(Peng et al. 2007). The initial sample consisted of nine sources.
Two of them are low persistent sources ending with seven burst-only sources (actually one was also detected
during outburst by RXTE-ASM). C02b carried out a follow-up of five of these 
sources, having the remaining two a ROSAT and a BeppoSAX a detected counterpart (C02b). 
One new source was recently detected by the BAT onboard Swift and it was followed up down to the 
likely level of quiescence (Wijnands et al. 2009).

\begin{table*}
\caption{Observational summary.}
\label{summary}
{\footnotesize 
\begin{center}
\begin{tabular}{ccccc}
Source             & Counterpart & Quiescent $L^+$ & Outburst & Peak $L^+$ \\
                   &             &(erg s$^{-1}$)   &          &(erg s$^{-1}$)\\
\hline
SAX J1324.5--6361  & ?           &$<7\times10^{33}$& N        & -- \\
SAX J1752.3--3138  & N           &  --             & N        & -- \\
SAX J1753.5--2349  & Y           &$\lsim1\times10^{33}$& Y    &$\gsim5\times10^{35}$\\
SAX J1806.5--2215  & ?           &$<4\times10^{33}$& Y        &$\gsim2\times10^{36}$ \\
SAX J1818.7+1424   & ?           &$<7\times10^{32}$& N        & -- \\
SAX J1828.5--1037  & Y           &$\lsim6\times10^{32}$& Y    &$\gsim1\times10^{36}$\\
SAX J2224.9+5421   & Y           &$\lsim8\times10^{32}$& N    & -- \\
Swift J1749.4--2807& Y           &$\lsim3\times10^{32}$& Y    &$\gsim1\times10^{36}$\\
\hline
Source                   & & Minimum $L$    & & Maximum $L$ \\
                         & & (erg s$^{-1}$) & &(erg s$^{-1}$)\\
\hline
RXS J170854.4--321857    & &$4\times10^{35}$& & $4\times10^{36}$\\  %60min 0.1 msun
RXS J171724.2--402934    & &$3\times10^{34}$& & $2\times10^{35}$\\  %10min <0.01msun
XMMU J174716.1--281048$^*$&&$2\times10^{34}$& & $2\times10^{35}$\\  %10min <0.01msun
\hline
\end{tabular}
\end{center}
}
\noindent $^+$ Unabsorbed luminosity in the 0.5--10 keV energy band. For quiescent luminosities we have 
upper limits since we have just an upper limit to the distance from Type I bursts. 

\noindent $^*$ Given the very high absorption we quote the unabsorbed luminosity in the 2--10 keV energy band.
\end{table*}

We observed the original set of seven sources with Swift XRT. We confirm and improve the detections in quiescence of 
SAX J1754 and SAX J2225, and discover the counterpart to SAX J1754. For SAX J1752 we have not enough data. 
For the other three sources we detect possible counterparts with different degrees of certainty. From this limited
sample we can hint that a common characteristic seems to be a hardness of the X--ray spectrum in quiescence.
This is at variance with classical neutron star transients in quiescence (Campana et al. 1998) but is more common in lower
luminosity transient in quiescence as accreting millisecond X--ray pulsars group. We also infer the presence of some 
variability in many of the proposed counterparts in quiescence, which likely is a common characteristic of neutron 
star transients in quiescence (Campana et al. 2004b).

In addition to burst-only sources we observed also three faint persistent (or quasi-persistent) sources with Swift XRT. 
These sources are characterized by strong variability (and RXS J171724 showed an outburst). 
These sources have softer spectra and higher `quiescent' fluxes with respect to burst-only sources.

The comparison between the two classes calls for an explanation. Faint persistent bursting sources  are difficult to 
explain within disk instability models. In our sample RXS J170854 has the highest mean persistent luminosity
(around $10^{36}\ergs$) and can be accounted for by a $\sim 60$ min binary system with a $\sim 0.1\msole$ companion.
Such a system can provide the required mean accretion rate (e.g. Bildsten \& Chakrabarty 2001; King 2000) and 
retain a stable accretion disk (e.g. Lasota et al. 2008). On the contrary RXS J171724 and XMMU J174716 have lower
mean luminosities ($\sim 5\times10^{34}\ergs$) and to maintain a stable system very short orbital periods ($\lsim 10$ min) 
{\it and} very low mass companions ($\lsim 0.01\msole$) are needed. These constraints seem to be too demanding and we
suggest that these systems are instead quasi-persistent systems (as suggested for XMMU J174716), i.e. with long outbursts like
classical transients as KS1731--260 or EXO 0748--676 but they will likely turn down to quiescence in the next years.

For burst-only sources we have detected three out of seven sources in outburst, likely indicating that these sources are really
transient in nature. Swift J1749, if typical, indicates that outbursts are faint and very short (from outburst to quiescence in 
less than a week), providing a justification for the  
Peak luminosities and mean accretion rates (for typical $\sim 10\%$ duty cycles) of burst-only sources are very low,
hinting to VFXTs. King \& Wijnands (2006) discussed several possibilities to explain the nature of VFXTs. The only one working 
with neutron star primaries concerns systems that formed with brown dwarfs (or even planetary systems) from the beginning or 
high inclination systems. We note here that the column densities we found for burst-only sources are consistent (or less) than 
the Galactic value thus constraining the presence of additional material.

\begin{acknowledgements}
I would like to thank R. Cornelisse for useful comments on an earlier version of the manuscript.
\end{acknowledgements}

\end{document}